\documentclass[superscriptaddress,secnumarabic,twocolumn,
 amssymb,amsmath,nobibnotes,aps,prd,showkeys,showpacs]{revtex4}
\usepackage{graphicx,epsfig,amssymb,amsmath,amstext}

\usepackage{graphicx}
\usepackage{bm}

\begin{document}

\title{Comment on ``Nonexistence of the final first integral in the Zipoy-Voorhees
 space-time''}

\author{Georgios Lukes-Gerakopoulos}
\email{gglukes@gmail.com}
\affiliation{Theoretical Physics Institute, University of Jena, 07743 Jena,
Germany}

\author{Andrzej J.~Maciejewski}
\email{maciejka@astro.ia.uz.zgora.pl}
\affiliation{J.~Kepler Institute of Astronomy, University of Zielona G\'ora,
Licealna 9, PL-65--417 Zielona G\'ora, Poland.}

\author{Maria Przybylska}
\email{M.Przybylska@proton.if.uz.zgora.pl}
\affiliation{Institute of Physics, University of Zielona G\'ora, Licealna 9,
65--417 Zielona G\'ora, Poland }

\author{Tomasz Stachowiak} \email{stachowiak@cft.edu.pl}
\affiliation{Center for Theoretical Physics PAS, Al. Lotnikow 32/46, 02-668
Warsaw, Poland }

\begin{abstract}

The accuracy of the numerical findings of \cite{LG12}, regarding the existence of
additional first integrals in the Zipoy-Voorhees space-time, was recently
questioned \cite{Maciejewski13}. In this comment, it is shown that the discrepancy
between the results of \cite{LG12} and \cite{Maciejewski13} is not due to issues
related to numerical accuracy, as claimed in \cite{Maciejewski13}, but due to
different choice of coordinates used in \cite{Maciejewski13}.

\end{abstract}

%
%
\maketitle

Numerical studies can be very effective in providing quantitative and qualitative
information about dynamical systems. Their accuracy can be checked quantitatively
and qualitatively as well. Namely, the adaptive step Cash-Karp Runge-Kutta used
in \cite{LG12} is an efficient tool for studying numerically the chaos in the 
Zipoy-Voorhees space-time. The accuracy of the numerical calculations of 
\cite{LG12} was discussed in the appendix of \cite{LG12}, where it was shown that
the relative errors in the calculations were below $10^{-12}$ (see Fig. 13 in
\cite{LG12}). This level of relative errors cannot justify the shift
between Fig. 5 of \cite{LG12} and Fig. 1 of \cite{Maciejewski13} as claimed in
\cite{Maciejewski13}, and in any case one should be very careful to discern 
any difference by just inspecting a surface of section like those
in \cite{LG12,Maciejewski13}. Furthermore, the detailed self-similar structures of
the islands of stability shown on the surface of section of \cite{LG12} is a
qualitative criterion which excludes the possibility of ``artificial'' chaos
produced by an inaccurate numerical scheme. 

While the algebraic analysis in \cite{Maciejewski13} is not influenced by this, the
comparison between Fig. 5 in \cite{LG12} and Fig. 1 of \cite{Maciejewski13} is
mistaken. The discrepancy between the two figures results from the very simple fact
that the prolate spheroidal coordinates $x,~y$ used in Fig. 1 of
\cite{Maciejewski13} are connected with the cylindrical $\rho,~z$ used in Figs.
of \cite{LG12} through the relations

\begin{equation}\label{eq:spcTcc}
 \rho=k\sqrt{(x^2-1)(1-y^2)},~~~z=k~x~y~~.
\end{equation}

\begin{figure}[h]
 \centerline{\includegraphics[width=0.44 \textwidth] {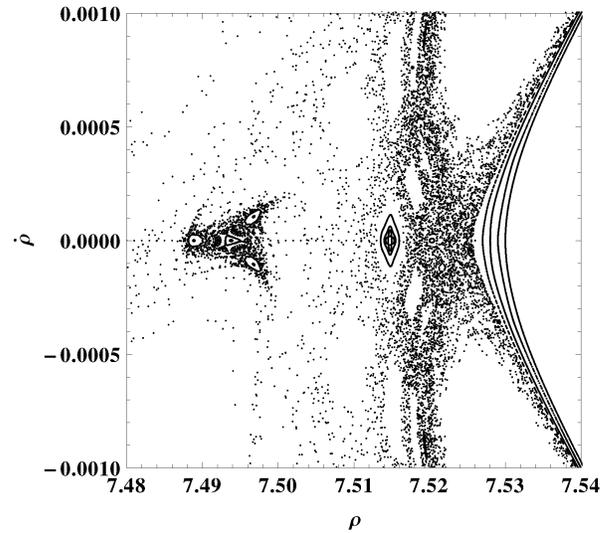}}
 \caption{ A detail from the surface of section $z=0$ ($\dot{z}>0$) for $\delta=2$,
 $p_0=-E=0.95$ and $L_z=3$.
 }
 \label{Example}
\end{figure}

The different coordinate systems causes Fig. \ref{Example} of this comment to be
``visibly shifted'' from Fig. 1 in \cite{Maciejewski13}, and one should use the 
eq. (\ref{eq:spcTcc}) in order to be able to compare the two figures, which was not
considered in \cite{Maciejewski13}.

It is easy to check the points mentioned above, and therefore the comments made in 
the introduction of \cite{Maciejewski13} about the quality of the numerical results
produced in \cite{LG12} are in fact too strong. Moreover, the analysis in
\cite{LG12} was not only based on the chaotic regions found around the main islands
of stability, but it was shown that the islands of the Birkhoff chains inside the
main islands indicate that the Zipoy-Voorhees is not an integrable system.

To summarize, the numerical findings in \cite{LG12} provide evidence for chaotic
dynamics, which in turn opens the question of integrability. This is where the
analytical proof of \cite{Maciejewski13} comes in, and the two studies can be
considered complementary to each other.  

\begin{acknowledgments}
 GL-G was supported by the DFG grant SFB/Transregio 7. For AM, MP and TS 
 this research has been supported by grant No. DEC-2011/02/A/ST1/00208 of National
 Science Centre of Poland.
\end{acknowledgments}

\end{document}